\documentclass[12pt]{article}
\pdfoutput=1

\usepackage{amsthm,amsfonts,amssymb,amsmath,dsfont}
\usepackage[pdftex]{graphicx,graphics,color}
\usepackage{color,colortbl}
\usepackage{lmodern}
\usepackage[english]{babel}
\definecolor{citecl}{rgb}{0.00,0.00,1.00}
\usepackage[colorlinks=true,linkcolor=black,citecolor=citecl,anchorcolor=blue,filecolor=blue,urlcolor=blue]{hyperref}
\usepackage[super,sort&compress]{natbib}
\usepackage[affil-it]{authblk}
\usepackage{subfigure}
\usepackage{hyphenat}

\DeclareMathAlphabet{\mathpzc}{OT1}{pzc}{m}{it}

\begin{document}


\newcommand{\nn}{\nonumber}
\newcommand{\bra}{\langle}
\newcommand{\ket}{\rangle}
\newcommand{\del}{\partial}
\newcommand{\vt}{\vec}
\newcommand{\dg}{^{\dag}}
\newcommand{\cg}{^{*}}
\newcommand{\T}{^{T}}
\newcommand{\vep}{\varepsilon}
\newcommand{\vth}{\vartheta}
\newcommand{\suml}{\sum\limits}
\newcommand{\prodl}{\prod\limits}
\newcommand{\intl}{\int\limits}
\newcommand{\til}{\tilde}
\newcommand{\wb}{\overline}
\newcommand{\mcl}{\mathcal}
\newcommand{\mfk}{\mathfrak}
\newcommand{\mds}{\mathds}
\newcommand{\mbb}{\mathbb}
\newcommand{\mrm}{\mathrm}
\newcommand{\mnl}{\mathnormal}
\newcommand{\ds}{\displaystyle}
\newcommand{\rmi}{\mathrm{i}}
\newcommand{\rme}{\mathrm{e}}
\newcommand{\rmd}{\mathrm{d}}
\newcommand{\rmD}{\mathrm{D}}
\newcommand{\vphi}{\varphi}
\newcommand{\stm}{\text{\textsf{s}}}
\newcommand{\bth}{\text{\textsf{b}}}
\newcommand{\itr}{\text{\textsf{i}}}
\renewcommand{\mod}{\mathrm{\,mod\,}}
\renewcommand{\b}{\bar}
\renewcommand{\dim}{\mbox{dim}}
\renewcommand{\-}{\,-}
\newcommand{\onlinecite}[1]{\hspace{-1 ex} \nocite{#1}\citenum{#1}} 
\newcommand{\malt}{\mathpzc}


\interfootnotelinepenalty=10000

\renewcommand{\topfraction}{0.85}
\renewcommand{\textfraction}{0.1}

\allowdisplaybreaks[1]

\setlength{\jot}{1ex}

\renewcommand{\theequation}{\thesection.\arabic{equation}}
\numberwithin{equation}{section}
\renewcommand{\thefigure}{\thesection.\arabic{figure}}
\numberwithin{figure}{section}
\renewcommand{\thefootnote}{\roman{footnote}}


\title{Beatification: Flattening the Poisson Bracket for Two-Dimensional Fluid and Plasma Theories}

\author[1]{Thiago F. Viscondi}
\author[1]{Iber\^e L. Caldas}
\author[2]{Philip J. Morrison}
\affil[1]{Institute of Physics, University of S\~ao Paulo, S\~ao Paulo, Brazil}
\affil[2]{Institute for Fusion Studies and Department of Physics, The University of Texas at Austin, Austin, TX 78712-1060, USA}

\date{}
\maketitle


\begin{abstract}
A perturbative method called \textit{beatification} is presented for a class 
of two-dimensional fluid and plasma theories. The Hamiltonian systems considered, 
namely the Euler, Vlasov-Poisson, Hasegawa-Mima, and modified Hasegawa-Mima equations, 
are naturally described in terms of noncanonical variables. The beatification procedure 
amounts to finding the correct transformation that removes the explicit variable dependence 
from a noncanonical Poisson bracket and replaces it with a fixed dependence on a chosen state 
in phase space. As such,  beatification is a major step toward casting the Hamiltonian system 
in its canonical form, thus enabling or facilitating the use of analytical and numerical 
techniques that require or favor a representation in terms of canonical, or beatified, 
Hamiltonian variables.
\end{abstract}


\section{Introduction}
\label{sec:introduction}

The Hamiltonian formalism is a fundamental pillar of theoretical physics, as the time evolution 
of every isolated system is expected to possess Hamiltonian structure. Although most literature 
on Hamiltonian systems uses canonical variables, numerous physical theories are Hamiltonian yet 
naturally described in terms of noncanonical variables,\footnote{The term \textit{noncanonical} 
is used to indicate a complete or overcomplete set of dynamical variables in terms of which 
a Hamiltonian system is not described by the canonical form of Hamilton's equations, but by a 
Poisson-bracket form with preserved algebraic properties. For further information on the fundamentals 
of noncanonical Hamiltonian systems we refer the reader to references~[\onlinecite{Morrison82b,Morrison98}].} 
such as many prominent fluid and plasma models,\cite{Morrison80a,Morrison82a,Morrison82b,Olver84a,Olver84b,
Olver86,Morrison98,Tassi09,Chandre10,Chandre14,D'Avignon16} the generalized coherent-state approach to 
semiclassical dynamics,\cite{Viscondi11a,Viscondi11b,Viscondi15,Viscondi16a} and even the time-dependent 
Schrödinger equation itself.\cite{Kramer81} 

For such noncanonical representations, the question then arises: How does one obtain
a global transformation from noncanonical to canonical variables in infinite-dimensional 
Hamiltonian systems, which are generally described by sets of partial differential equations?
In the present paper, this problem is partially addressed by using an analytical method known 
as \textit{beatification}.\cite{Morrison16,Viscondi16b} 

Beatification is a perturbative procedure through which the explicit variable dependence 
of a noncanonical Poisson bracket is replaced by a fixed dependence on a chosen state in 
phase space, designated as the \textit{reference state}. As a result of the beatifying transformation, 
the Hamiltonian functional undergoes an increase in its degree of nonlinearity, so that the removal 
of the dynamical variable in the Poisson bracket is compensated.\cite{Morrison16,Viscondi16b} Another 
important consequence of beatification is that it greatly facilitates the search for canonical variables, 
as can be readily verified in the case of a finite-dimensional Hamiltonian system.\cite{Morrison16,Viscondi16b} 
Therefore, beatification can be seen as a preparatory step toward canonization, as implied by its name.

Considering both finite- and infinite-dimensional Hamiltonian systems, the beatification method was first presented 
in reference~[\onlinecite{Morrison16}], where the beatifying transformation was derived, to its lowest perturbative 
order, by using an equilibrium as reference state. In a subsequent work,\cite{Viscondi16b} considering specifically 
the Poisson bracket for vorticity-like variables in two-dimensional fluid and plasma theories, beatification was 
extended to second order and generalized to take into account completely arbitrary reference states. In the present 
paper, considering again the case of two-dimensional fluid and plasma models, the beatifying transformation is 
further extended to infinite perturbative order, also termed \textit{complete beatification}. In this way, an 
appropriate setting is established for an in-depth investigation of some important Hamiltonian models, namely 
the two-dimensional Euler equation,\cite{Morrison98} the standard\cite{Charney71,Hasegawa78,Tassi09,Chandre10} 
and modified\cite{Smolyakov00,Chandre14} versions of the Hasegawa-Mima equation, and the one-degree-of-freedom 
Vlasov-Poisson equation.\cite{Morrison80b,Ye91} We mention that, in essence, the equivalent of complete beatification 
was obtained for the special case of expansion about shear flow equilibria in reference~[\onlinecite{Zakharov88}] 
and similar ideas were explored for the Vlasov-Poisson system by introduction of a generating function in 
reference~[\onlinecite{Ye91}].

A beatified Hamiltonian system exhibits some significant analytical advantages over its original noncanonical 
form. First, due to the removal of the explicit variable dependence from the Poisson bracket, the beatified 
system can have its degrees of freedom truncated without loss of Hamiltonian structure.\cite{Viscondi16b} 
Such a reduction process is very useful for obtaining a finite set of dynamical equations from an 
infinite-dimensional system, while retaining the Hamiltonian properties of the latter. Potential 
uses for this Hamiltonian truncation procedure include constructing low-dimensional models for 
describing specific physical mechanisms\cite{Sugihara68,Karplyuk73,Wersinger80,Turner80,Verheest82,
Romeiras83,Kueny95a,Kueny95b,Lopes96,Chian96,Pakter97,Chen00,LashmoreDavies01,LashmoreDavies05,
Kolesnikov05a,Kolesnikov05b,Benzekri05,Connaughton10} and obtaining semi-discrete schemes for 
numerical integration of partial differential equations, as an alternative to techniques used 
or derived, for example, in references~[\onlinecite{Arakawa66,Arakawa81,Salmon89,Cockburn91,
Cheng13,Zhou14,Xiao16,Kraus16}]. Second, as a collateral effect of beatification, all Casimir 
invariants\footnote{Casimir invariants are constants of motion arising from degeneracy of a 
noncanonical Poisson Bracket.} of a system become linear in the dynamical variables. As an evident 
consequence, analytical manipulation of these constants of motion is greatly simplified. Third, as 
a perturbative approach, beatification can be used to simplify or emphasize the local dynamics about 
a particular phase-space point, chosen as reference state.

The remainder of this paper is organized as follows. Section~\ref{sec:hamiltform} briefly introduces 
the Hamiltonian formalism for two-dimensional fluid and plasma theories. In this context, the pertinent 
Poisson brackets, Hamiltonian functionals, and equations of motion are presented. In section~\ref{sec:casimir}, 
the Casimir invariants of the relevant theories are derived. Section~\ref{sec:beatification} presents the 
beatification procedure itself, which is composed of two key steps. First, the field variable is rewritten 
as the sum of a reference state and a perturbative field. Second, an additional transformation is imposed 
on the perturbative field, so that the explicit variable dependence is eliminated in the Poisson bracket. 
At the end of section~\ref{sec:beatification}, the effect of the beatifying transformation on the Casimir 
invariants is examined in detail. Section~\ref{sec:invtrans} analyzes the particular case of a finite-order 
beatification procedure and presents a recurrence formula for the inverse of the beatifying transformation.
In section~\ref{sec:conclusion}, the main findings of the paper are summarized and potential applications 
are discussed. Finally, in appendix~\ref{app:brackettrans}, the suppression of the variable dependence 
in the Poisson bracket as a result of the beatifying transformation is demonstrated.

\section{Hamiltonian formulation}
\label{sec:hamiltform}

In this section, we present the main elements of the Hamiltonian formalism for four important 
fluid and plasma models, namely the two-dimensional Euler equation, the standard and modified 
versions of the Hasegawa-Mima equation, and the one-degree-of-freedom Vlasov-Poisson equation. 
As a first step, we introduce the fundamental Poisson bracket that connects all the Hamiltonian 
systems considered:\footnote{The proof that the antisymmetric bilinear operation~\eqref{eq:2.1} 
is indeed a Poisson bracket, that is, it satisfies the Jacobi identity, is given in 
references~[\onlinecite{Morrison81,Morrison82b}].}
\begin{equation}
 \left\{F,G\right\}=\intl_{\mcl{D}}\rmd^{2}r\,\frac{\delta F}{\delta\omega}
 \mcl{J}(\omega)\frac{\delta G}{\delta\omega},
 \label{eq:2.1}
\end{equation}
\noindent where $\omega=\omega(x,y;t)$ is a vorticity-like scalar field on the two-dimensional 
domain~$\mcl{D}$, $\rmd^{2}r=\rmd{x}\rmd{y}$, $F$ and $G$ are two arbitrary functionals of 
$\omega$, and $\mcl{J}$ is the \textit{Poisson operator}, which is defined as: 
\begin{equation}
 \mcl{J}(f)g=-[f,g],
 \label{eq:2.2}
\end{equation}
\noindent for two arbitrary functions $f$ and $g$ on the domain~$\mcl{D}$ and 
$[f,g]=(\del_{x}f)(\del_{y}g)-(\del_{y}f)(\del_{x}g)$. Notice that, except for a minus sign, 
the Poisson operator is just the Jacobian determinant~$\del(f,g)/\del(x,y)$ or, equivalently, 
the $z$ component of the cross product between gradients, that is, $(\nabla f\times \nabla g)_{z}$. 
Also presented in equation~\eqref{eq:2.1}, the functional derivatives are defined as usual:
\begin{equation}
 \delta F[\omega;\delta\omega]=\left.\frac{\rmd}{\rmd\vep}F[\omega+\vep\delta\omega]\right|_{\vep=0}
 =\intl_{\mcl{D}}\rmd^{2}r\,\frac{\delta F}{\delta\omega}\delta\omega.
 \label{eq:2.3}
\end{equation}

For simplicity, from now on, we choose the domain~$\mcl{D}$ as a normalized 2-torus, so that
$x,y\in[0,1)$ and periodic boundary conditions are implied. This choice allows us to promptly 
get rid of boundary terms whenever an integration by parts over $\mcl{D}$ is performed.

Once the Poisson brackets of the relevant systems are known, in order to conclude the presentation 
of their Hamiltonian formalism, we also have to establish their Hamiltonian functionals. As we shall 
see shortly, the required functionals are contained in the following general formula:\cite{Morrison03}
\begin{equation}
 H[\omega]=\intl_{\mcl{D}}\rmd^{2}r\, 
 h_{1}(r)\omega(r;t)
 +\frac{1}{2}\intl_{\mcl{D}}\rmd^{2}r\intl_{\mcl{D}}\rmd^{2}r'\,
 \omega(r;t)h_{2}(r,r')\omega(r';t),
 \label{eq:2.9}
\end{equation}
\noindent for $r=(x,y)$. The quantities $h_{1}(r)$ and $h_{2}(r,r')$ of equation~\eqref{eq:2.9}
describe, respectively, the free-motion and two-point-interaction energies of the system. For 
the Vlasov-Poisson equation, considering particles with normalized mass and charge, the Hamiltonian 
functional is given with $h_{1}(r)=y^{2}/2$ and $h_{2}(r,r')=V(|x-x'|)$, where $V(|x-x'|)$ is the 
Green's function for the Poisson equation. In the cases of the Euler, Hasegawa-Mima, and modified 
Hasegawa-Mima equations, the Hamiltonian functionals are jointly specified by the expressions 
$h_{1}(r)=\mcl{L}^{-1}\lambda$ and $h_{2}(r,r')=-\delta^{(2)}(r-r')\mcl{L}^{-1}$, where $\lambda=\lambda(x,y)$ 
is a specified function on the domain~$\mcl{D}$ and $\mcl{L}$ is a linear operator, which is also required 
to be self-adjoint with respect to the following scalar product between functions on the domain~$\mcl{D}$:
\begin{equation}
 \bra f,g\ket=\intl_{\mcl{D}}fg\,\rmd^{2}r.
 \label{eq:2.6}
\end{equation}

In each of the Hamiltonian models considered in this paper, the quantities $\omega$, 
$\lambda$ and $\mcl{L}$ perform different roles. In the case of the two-dimensional 
Euler equation, $\omega$ stands for the usual scalar vorticity, $\lambda=0$, and $\mcl{L}$ 
is the two-dimensional Laplacian operator, that is, $\mcl{L}=\Delta=\del_{x}^{2}+\del_{y}^{2}$. 
For the standard form of the Hasegawa-Mima equation, $\omega$ is a vorticity-like field related 
to the electrostatic potential~$\phi$ by the transformation~$\omega=\mcl{L}\phi+\lambda$, $\lambda$ 
is a function depending on the electron density at equilibrium,\cite{Chandre10} and $\mcl{L}=\Delta-1$. 
For the modified Hasegawa-Mima equation, $\omega$ and $\lambda$ have the same meanings as for the standard 
version, while $\mcl{L}=\Delta-1+\mcl{P}$, where the operator~$\mcl{P}$ denotes integration over the $y$-axis, 
that is, $\mcl{P}f=\int_{0}^{1}f{\rmd}y$ for any function~$f$ on the domain~$\mcl{D}$.\footnote{The self-adjointness 
of the Laplacian operator~$\Delta$ follows directly from integration by parts under periodic boundary conditions. 
Similarly, we can readily demonstrate that the projector~$\mcl{P}$ is also a self-adjoint operator by performing 
the Fourier transform of both arguments of the pertinent scalar product.} For the one-degree-of-freedom Vlasov-Poisson 
equation, $\omega$ is the phase-space probability distribution of a one-species plasma and the quantities $\lambda$ 
and $\mcl{L}$ are not defined.

A distinguishing feature in the description of the one-degree-of-freedom Vlasov-Poisson 
equation is that the two-dimensional domain~$\mcl{D}$ stands for the phase space of a 
charged particle restricted to a one-dimensional configuration space, unlike the three 
other systems considered, in which $\mcl{D}$ is the actual space occupied by the fluid 
or plasma. In other words, for the Euler, Hasegawa-Mima, and modified Hasegawa-Mima 
equations, the coordinates $x$ and $y$ denote the spatial position of a fluid or 
plasma infinitesimal element, whereas, in the case of the Vlasov-Poisson equation, 
$x$ and $y$ represent, respectively, the position and linear momentum variables 
of a phase-space probability density function.

For completeness, we now work out the equations of motion for the four models considered 
here. To this end, we write down the usual Hamiltonian relation between the time variation 
of the field~$\omega$ and the Poisson bracket~\eqref{eq:2.1}:
\begin{equation}
 \frac{\del\omega}{\del t}=\{\omega,H\}
 =\mcl{J}(\omega)\frac{\delta H}{\delta\omega}.
 \label{eq:2.10}
\end{equation}

From equation~\eqref{eq:2.9}, we can readily calculate the derivative of the Hamiltonian functional, 
which is found to be $\delta H/\delta\omega=h_{1}(r)+\int_{\mcl{D}}\rmd^{2}r'\,h_{2}(r,r')\omega(r';t)$. 
By substituting this result into identity~\eqref{eq:2.10} with the appropriate values for $h_{1}$ and 
$h_{2}$, we first obtain a general expression for the Euler, Hasegawa-Mima, and modified Hasegawa-Mima 
equations:
\begin{equation}
 \frac{\del\omega}{\del t}=[\omega,\mcl{L}^{-1}(\omega-\lambda)].
 \label{eq:2.11}
\end{equation}

Then, by considering the suitable choices for $h_{1}$ and $h_{2}$, 
we derive the one-degree-of-freedom Vlasov-Poisson equation:
\begin{equation}
 \frac{\del\omega}{\del t}+y\frac{\del\omega}{\del x}-\frac{\del\phi}{\del x}\frac{\del\omega}{\del y}=0,
 \label{eq:2.12}
\end{equation}
\noindent in which $\phi(x)=\int_{\mcl{D}}V(|x-x'|)\omega(r')\rmd^{2}r'$ is the electrostatic potential.

\section{Casimir invariants}
\label{sec:casimir}

The Casimir invariants associated with a certain Poisson bracket are defined as the quantities 
whose functional derivatives belong to the null space of the corresponding Poisson operator. 
Therefore, in the case of the bracket~\eqref{eq:2.1}, the Casimir invariants are determined 
by the following identity:
\begin{equation}
 \mcl{J}(\omega)\frac{\delta\mnl{C}}{\delta\omega}=0.
 \label{eq:3.1}
\end{equation}

According to equations \eqref{eq:2.1} and \eqref{eq:3.1}, note that the Poisson bracket between 
a Casimir invariant~$\mnl{C}[\omega]$ and an arbitrary functional~$F[\omega]$ is identically zero, 
that is, $\{F,\mnl{C}\}=0$ for any functional~$F$ of the field~$\omega$. As a direct consequence, 
the Casimir invariants are constants of motion for any choice of Hamiltonian functional:
\begin{equation}
 \frac{\rmd\mnl{C}}{\rmd t}=\{\mnl{C},H\}=0.
 \label{eq:3.2}
\end{equation}

By employing definition~\eqref{eq:2.2}, we can readily demonstrate 
that the Poisson operator~$\mcl{J}$ satisfies the following equation: 
\begin{equation}
 \mcl{J}(\omega)\malt{g}(\omega)=0,
 \label{eq:3.3}
\end{equation}
\noindent where $\malt{g}(\omega)$ denotes an arbitrary function of 
the vorticity-like field~$\omega$. By comparing identities \eqref{eq:3.1} 
and \eqref{eq:3.3} for $\malt{g}(\omega)=\rmd\malt{f}(\omega)/\rmd\omega$, 
we conclude that the Casimir invariants for the Poisson bracket~\eqref{eq:2.1} 
must take the form
\begin{equation}
 \mnl{C}[\omega]=\intl_{\mcl{D}}\malt{f}(\omega)\,\rmd^{2}r,
 \label{eq:3.4}
\end{equation}
\noindent as the functional derivative of the above expression is simply 
given by $\delta\mnl{C}/\delta\omega=\rmd\malt{f}(\omega)/\rmd\omega$.

As a concluding remark to this section, we observe that, since the null space of a canonical 
Poisson operator is trivial, the existence of Casimir invariants is an exclusive property 
of noncanonical Hamiltonian systems. 

\section{Beatification}
\label{sec:beatification}

Note that, excluding the possible dependence on the field~$\omega$ arising from the derivatives of the 
functionals $F$ and $G$, the Poisson Bracket~\eqref{eq:2.1} has its own dependence on the field variable, 
which is contained in the Poisson operator~$\mcl{J}(\omega)$. The removal of this explicit variable dependence 
constitutes the primary purpose of the perturbative method known as \textit{beatification}. In this section, 
we present the main result of the paper, namely, the infinite-order beatifying transformation for the Poisson 
bracket~\eqref{eq:2.1}.

The beatification procedure is composed of two stages. First, the field~$\omega$ is rewritten as the sum 
of a reference state~$\omega_{0}$ and a perturbative variable~$\mu$. Second, a nonlinear change of variables 
is performed on $\mu$, which is recast in terms of a new perturbative field~$\eta$. As a consequence of 
this second transformation, the Poisson operator becomes independent of the field variable.

The first step in the beatification procedure amounts to the following shift of the field~$\omega$:
\begin{equation}
 \omega(x,y;t)=\omega_{0}(x,y)+\vep\mu(x,y;t).                
 \label{eq:4.1}
\end{equation}
Here the quantity~$\omega_{0}$, designated as the \textit{reference state}, is an arbitrary time-independent 
function on the domain~$\mcl{D}$. In general, a specific choice for the state~$\omega_{0}$ is determined by 
physical or mathematical features we want to introduce or emphasize in the Hamiltonian system under study. 
In equation~\eqref{eq:4.1}, we have also presented the new dynamical variable~$\mu$ and the perturbative 
parameter~$\vep$, which is used to keep track of terms from different perturbative orders during the 
beatification process. 

As a preparation for our future developments, it is necessary to recast the Poisson 
bracket \eqref{eq:2.1} in terms of the variable~$\mu$. For this purpose, we shall 
make use of the \textit{chain rule for functional derivatives}:\cite{Morrison98}
\begin{equation}
 \frac{\delta F}{\delta\chi}=
 {\left(\frac{\delta\xi}{\delta\chi}\right)\!}\dg
 \frac{\delta F}{\delta\xi},
 \label{eq:2.5}
\end{equation}
\noindent where $\chi$ and $\xi$ are two fields variables related by the transformation 
$\xi=\xi\{\chi\}$.\footnote{The curly-bracket notation indicates that $\xi$ depends 
on $\chi$ in an arbitrary manner, which may include, for example, derivatives of the 
latter.} The quantity $\delta\xi/\delta\chi$ denotes the linear operator that, when 
applied on the variation~$\delta\chi$, results in the corresponding variation~$\delta\xi$. 
The operator~$(\delta\xi/\delta\chi)\dg$ symbolizes the adjoint of $\delta\xi/\delta\chi$
with respect to the scalar product~\eqref{eq:2.6}.

By applying the chain rule~\eqref{eq:2.5}, we obtain that the functional derivatives with respect 
to the fields $\omega$ and $\mu$ are related by $\delta F/\delta\omega=\vep^{-1}(\delta F/\delta\mu)$. 
Upon substitution of this result into equation~\eqref{eq:2.1}, a new form for the Poisson bracket 
is achieved: 
\begin{equation}
 \vep^{2}\{F,G\}=\intl_{\mcl{D}}\rmd^{2}r
 \frac{\delta F}{\delta\mu}
 [\mcl{J}(\omega_{0})+\vep\mcl{J}({\mu})]
 \frac{\delta G}{\delta\mu}.            
 \label{eq:4.3}
\end{equation}

As we can see in the above equation, transformation~\eqref{eq:4.1} divided 
the original Poisson operator~$\mcl{J}(\omega)$ into two terms. The first one 
is simply the operator~$\mcl{J}(\omega)$ calculated at the reference state, that 
is, $\mcl{J}(\omega_{0})$. The second term, given by $\mcl{K}(\mu)={\vep}\mcl{J}(\mu)$, 
constitutes the perturbative part of the Poisson operator and is now responsible for 
all explicit variable dependence of the Poisson bracket. 

The second step of the beatification procedure corresponds to finding a transformation~$\eta=\eta\{\mu\}$ which 
eliminates the term~$\mcl{K}(\mu)$ in equation~\eqref{eq:4.3} while preserving the operator~$\mcl{J}(\omega_{0})$.
In this way, the ultimate goal of beatification is attained, namely, removing the dependence on the dynamical 
field from the Poisson operator~$\mcl{J}(\omega)$ and replacing it with the reference state~$\omega_{0}$.

According to the functional chain rule, presented in equation~\eqref{eq:2.5}, 
the derivatives with respect the perturbative variable~$\mu$ and the beatified 
field~$\eta$ are related by $\delta F/\delta\mu=(\delta\eta/\delta\mu)\dg\delta F/\delta\eta$, 
in which $\delta\eta/\delta\mu$ is the linear operator that transforms an infinitesimal 
variation~$\delta\mu$ into a corresponding variation~$\delta\eta$. By applying the chain 
rule to both functional derivatives in equation~\eqref{eq:4.3}, we rewrite the Poisson 
bracket in terms of the field~$\eta$:
\begin{equation}
 \vep^{2}\{F,G\}=\intl_{\mcl{D}}\rmd^{2}r
 \frac{\delta F}{\delta\eta}\,\til{\mcl{J}}\,\frac{\delta G}{\delta\eta},
 \label{eq:4.4}
\end{equation}
\noindent where the transformed Poisson operator is given by 
\begin{equation}
 \til{\mcl{J}}=\frac{\delta\eta}{\delta\mu}
 [\mcl{J}(\omega_{0})+\vep\mcl{J}(\mu)]
 {\left(\frac{\delta\eta}{\delta\mu}\right)\!}\dg,
 \label{eq:4.5}
\end{equation}
\noindent for $\mu=\mu\{\eta\}$. As previously mentioned, the beatifying transformation~$\eta=\eta\{\mu\}$ 
is found by demanding that the right-hand side of equation~\eqref{eq:4.5} be reduced to the value of 
the Poisson operator~$\mcl{J}(\omega)$ at the reference state. In short, the beatifying transformation 
is defined by the following identity: 
\begin{equation}
 \til{\mcl{J}}=\mcl{J}(\omega_{0}).
 \label{eq:4.6}
\end{equation}

By employing the above equation, the beatifying transformation can be derived through an order-by-order 
perturbative process, whose zeroth-order term is taken as the identity transformation.\cite{Morrison16}
In the present paper, we do not follow this approach. Instead, we simply propose an expression 
for the beatifying transformation and then prove that it indeed satisfies equation~\eqref{eq:4.6}.
In accordance with this plan of action, we now present the infinite-order beatifying transformation 
for the Poisson bracket~\eqref{eq:2.1}:
\begin{equation}
 \eta=\suml_{j=0}^{\infty}\frac{\vep^{j}}{(j+1)!}\mcl{B}^{j}\mu^{j+1},
 \label{eq:4.7}
\end{equation}
\noindent where, for notational simplicity, we have also defined an auxiliary operator:
\begin{equation}
 \mcl{B}f=-\frac{1}{2}\left(\del_{x}\frac{f}{\del_{x}\omega_{0}}
 +\del_{y}\frac{f}{\del_{y}\omega_{0}}\right),
 \label{eq:4.8}
\end{equation}
\noindent for any function~$f$ on the domain~$\mcl{D}$.\footnote{In the particular 
case where the state~$\omega_{0}$ depends on only one coordinate, the definition of 
the operator~$\mcl{B}$ needs to be slightly modified. For $\omega_{0}=\omega_{0}(x)$, 
the correct expression is given by $\mcl{B}f=-\del_{x}(f/\del_{x}\omega_{0})$. 
The corresponding expression for $\omega_{0}=\omega_{0}(y)$ is evident.} As 
expected from such perturbative expansion, expression~\eqref{eq:4.7} constitutes 
a near-identity transformation, that is, the perturbative series approaches the 
identity transformation~$\eta=\mu$ as the parameter~$\vep$ goes to zero.

Due to its relatively high complexity and length, the proof that transformation~\eqref{eq:4.7} 
satisfies identity~\eqref{eq:4.6} is left to appendix~\ref{app:brackettrans}. In the remainder 
of this section, we shall discuss the Casimir invariants of the beatified Poisson operator 
and their relation to the original Casimir functionals, presented in equation~\eqref{eq:3.4}.

Analogously to equation~\eqref{eq:3.1}, the Casimir invariants associated 
with the Poisson operator~\eqref{eq:4.6} are defined by the following 
identity:
\begin{equation}
 \mcl{J}(\omega_{0})\frac{\delta\til{\mnl{C}}[\eta]}{\delta\eta}=0,
 \label{eq:4.9}
\end{equation}
\noindent where we have introduced the tilde notation to specifically denote the Casimir 
functionals of the beatified field, that is, $\til{\mnl{C}}=\til{\mnl{C}}[\eta]$. By 
employing definition~\eqref{eq:2.2} in a completely similar way to equation~\eqref{eq:3.3}, 
we can readily show that
\begin{equation}
 \mcl{J}(\omega_{0})\malt{g}(\omega_{0})=0,
 \label{eq:4.10}
\end{equation}
\noindent for any function~$\malt{g}$ of the state~$\omega_{0}$. By comparing equations 
\eqref{eq:4.9} and \eqref{eq:4.10}, we conclude that the Casimir invariants~$\til{\mnl{C}}[\eta]$ 
must be linear in the field~$\eta$. For this reason, we define the following general expression 
for the beatified Casimir functionals:
\begin{equation}
 \til{\mnl{C}}[\eta]=\intl_{\mcl{D}}\rmd^{2}r
 \left[\malt{f}(\omega_{0})+{\vep}\frac{\rmd\malt{f}(\omega_{0})}{\rmd\omega_{0}}\eta\right],
 \label{eq:4.11}
\end{equation}
\noindent in which $\malt{f}$ corresponds to a second arbitrary function of $\omega_{0}$.
For the purpose of demonstrating the validity of the above definition, we present
the functional derivative of this expression:
\begin{equation}
 \frac{\delta\til{\mnl{C}}[\eta]}{\delta\eta}={\vep}\frac{\rmd\malt{f}(\omega_{0})}{\rmd\omega_{0}}.
 \label{eq:4.12}
\end{equation}

As a direct consequence of equations \eqref{eq:4.10} and \eqref{eq:4.12} 
for $\malt{g}(\omega_{0})=\vep[\rmd\malt{f}(\omega_{0})/\rmd\omega_{0}]$, 
we observe that definition~\eqref{eq:4.11} indeed satisfies 
identity~\eqref{eq:4.9}.

Equations \eqref{eq:3.4} and \eqref{eq:4.11} provide general expressions 
for the Casimir invariants respectively associated with the Poisson operators 
$\mcl{J}(\omega)$ and $\mcl{J}(\omega_{0})$. However, despite knowing the 
transformations connecting the fields $\omega$ and $\eta$, the direct relation 
between the functionals $\mnl{C}[\omega]$ and $\til{\mnl{C}}[\eta]$ has not 
yet been identified. In order to precisely relate these two equivalent sets 
of dynamical invariants, we now substitute transformation~\eqref{eq:4.7} into 
definition~\eqref{eq:4.11}: 
\begin{equation}
 \til{\mnl{C}}[\eta]=\intl_{\mcl{D}}\rmd^{2}r
 \left\{\malt{f}(\omega_{0})
 +\suml_{j=0}^{\infty}\frac{\vep^{j+1}}{(j+1)!}
 \left[(\mcl{B}\dg)^{j}\frac{\rmd\malt{f}(\omega_{0})}{\rmd\omega_{0}}\right]
 \mu^{j+1}\right\}.
 \label{eq:4.13}
\end{equation}

In the above equation, we have introduced the adjoint of operator~\eqref{eq:4.8}, 
which is explicitly given by
\begin{equation}
 \mcl{B}\dg f=\frac{1}{2}\left(\frac{\del_{x}f}{\del_{x}\omega_{0}}
 +\frac{\del_{y}f}{\del_{y}\omega_{0}}\right),
 \label{eq:4.14}
\end{equation}
\noindent for any function~$f$ on the domain~$\mcl{D}$. Equation~\eqref{eq:4.13} can be greatly 
simplified with the aid of an important property of the operator~$\mcl{B}\dg$: 
\begin{equation}
 \mcl{B}\dg\malt{g}(\omega_{0})=\frac{\rmd\malt{g}(\omega_{0})}{\rmd\omega_{0}},
 \label{eq:4.15}
\end{equation}
\noindent which follows directly from identity~\eqref{eq:4.14} for any function $\malt{g}$ of the 
state~$\omega_{0}$. By making successive uses of property~\eqref{eq:4.15}, equation~\eqref{eq:4.13}
reduces to
\begin{equation}
 \begin{aligned}
 \til{\mnl{C}}[\eta]
 &=\intl_{\mcl{D}}\rmd^{2}r\left[\malt{f}(\omega_{0})
 +\suml_{j=1}^{\infty}\frac{\vep^{j}}{j!}
 \frac{\rmd^{j}\malt{f}(\omega_{0})}{\rmd\omega_{0}^{j}}
 {\mu}^{j}\right]\\
 &=\intl_{\mcl{D}}\malt{f}(\omega_{0}+\vep\mu)\,\rmd^{2}r
 =\intl_{\mcl{D}}\malt{f}(\omega)\,\rmd^{2}r
 =\mnl{C}[\omega].
 \end{aligned} 
 \label{eq:4.16}
\end{equation}

The transition from the first to the second line of the above equation has been accomplished 
by identifying the Taylor series of the function~$\malt{f}(\omega_{0}+\vep{\mu})$. As evidenced 
by identity~\eqref{eq:4.16}, the function~$\malt{f}$ found in definition~\eqref{eq:4.11} is exactly 
the same as that of equation~\eqref{eq:3.4}. In this way, the effect of the infinite-order beatifying 
transformation on the Casimir invariants becomes completely known.

\section{Finite-order beatification and inverse transformation}
\label{sec:invtrans}

In many situations of practical interest, such as the derivation of low-dimensional Hamiltonian 
models,\cite{Morrison16,Viscondi16b} the beatification procedure is more appropriately used as 
a finite-order perturbative method. In this case, the beatifying transformation~\eqref{eq:4.7} 
is truncated at a predetermined power of the parameter~$\vep$:
\begin{equation}
 \eta^{(n)}\{\mu\}=\suml_{j=0}^{n}\frac{\vep^{j}}{(j+1)!}\mcl{B}^{j}\mu^{j+1},
 \label{eq:5.1}
\end{equation}
\noindent where $\eta^{(n)}$ is the $n$-th order beatified field. By using the above equation instead 
of the complete transformation~\eqref{eq:4.7}, the Poisson operator~\eqref{eq:4.5} takes the following 
form:
\begin{equation}
 \til{\mcl{J}}=\mcl{J}(\omega_{0})+O(\vep^{n+1}),
 \label{eq:5.2}
\end{equation}
\noindent that is, the beatified Poisson operator of equation~\eqref{eq:4.6} 
is again obtained, but this time the result holds only up to a certain perturbative 
order. The proof that the finite-order beatifying transformation provides the Poisson 
operator~\eqref{eq:5.2} is obtained by retracing the steps of appendix~\ref{app:brackettrans}, 
where the proof for the complete beatification is given.

As a consequence of identity~\eqref{eq:5.2}, the Casimir functionals~$\til{\mnl{C}}[\eta^{(n)}]$ 
must have leading order terms, excluding constants, which are linear in the field~$\eta^{(n)}$, 
in complete analogy with equation~\eqref{eq:4.11}. Therefore, by repeating the reasoning of 
section~\ref{sec:beatification} with the finite-order transformation~\eqref{eq:5.1}, we can 
show that the original Casimir invariants~$\mnl{C}[\omega]$, presented in equation~\eqref{eq:3.4}, 
are related to the following beatified functionals:
\begin{equation}
 \til{\mnl{C}}[\eta^{(n)}]=\intl_{\mcl{D}}\rmd^{2}r\left[\malt{f}(\omega_{0})
 +\vep\frac{\rmd\malt{f}(\omega_{0})}{\rmd\omega_{0}}\eta^{(n)}\right]
 +O(\vep^{n+2}).
 \label{eq:5.3}
\end{equation}

According to the above expression, the Casimir invariants~$\mnl{C}[\omega]$ 
are also linearized by the finite-order beatifying transformation, but only 
if the terms of order~$\vep^{n+2}$ and higher are discarded. 

An actual application of the finite-order beatification procedure to 
the investigation of a specific dynamical system often relies on the 
knowledge of the inverse of transformation~\eqref{eq:5.1}. Among other 
possible purposes, the inverse transformation is particularly necessary
in calculating the beatified form of Hamiltonian functionals.

The inverse of the finite-order beatifying transformation can be obtained through 
a recursive order-by-order process. That is, given that the $j$-th order inverse 
transformation~$\mu^{(j)}=\mu^{(j)}\{\eta\}$ is known for $j=0,1,\ldots,(n-1)$, 
the $n$-th order inverse transformation is determined by the following recurrence 
relation:
\begin{equation}
 \mu^{(n)}\{\eta\}=\eta-\suml_{j=1}^{n}\frac{\vep^{j}}{(j+1)!}
 \mcl{B}^{j}\left[\mu^{(n-j)}\{\eta\}\right]^{j+1},
 \label{eq:5.4}
\end{equation}
\noindent where it is implied that $\mu^{(0)}\{\eta\}=\eta$. Note that 
the exponentiation of $\mu^{(n-j)}\{\eta\}$ on the right-hand side of 
equation~\eqref{eq:5.4} can give rise to terms of order greater than
$\vep^{n}$. These spurious terms must be discarded so that the recurrence 
formula produces a consistent result for $\mu^{(n)}\{\eta\}$.

Notice that, upon substitution of expression~\eqref{eq:4.4} for $n>0$, a Hamiltonian 
functional~$H[\omega]=H[\omega_{0}+\vep\mu]$ experiences an increase in its degree of 
nonlinearity. This effect constitutes a compensation for the suppression of the variable 
dependence in the Poisson bracket, so that the corresponding equations of motion are not 
subjected to a decrease in their degree of nonlinearity.

\section{Conclusion}
\label{sec:conclusion}

Beatification is a perturbative method with the primary purpose of eliminating
the variable dependence of a noncanonical Poisson operator by replacing it with a chosen 
reference state. As the main result of this paper, we present the infinite-order beatifying 
transformation for the fundamental Poisson bracket of four important fluid and plasma 
Hamiltonian models, namely, the two-dimensional Euler equation, the standard and modified 
versions of the Hasegawa-Mima equation, and the one-degree-of-freedom Vlasov-Poisson equation.
This work builds on previous studies\cite{Morrison16,Viscondi16b} by extending the beatification 
procedure to infinite perturbative order.

The noncanonical Hamiltonian formalism for two-dimensional fluid and plasma theories was briefly 
outlined in section~\ref{sec:hamiltform}. Although this discussion was focused on four particular 
models, we would like to point out that the applicability of our central results is not restricted 
to the dynamical systems explicitly considered, as the Poisson bracket~\eqref{eq:2.1} takes part 
in the Hamiltonian description of a fairly broad class of continuous media theories. As clearly 
stated by the defining relation~\eqref{eq:4.6}, the beatification procedure is independent of 
the specific choice for the Hamiltonian functional, since it depends only on the form of the 
Poisson operator. Consequently, the main results of section~\ref{sec:beatification}, such as 
the beatifying transformation~\eqref{eq:4.7} and the beatified Casimir invariants~\eqref{eq:4.11}, 
have geometrical nature and are not affected by purely dynamical aspects of a particular Hamiltonian 
system.

An interesting secondary effect of beatification is the linearization of the Casimir functionals,
which is a direct consequence of definition~\eqref{eq:4.6}. At the end of section~\ref{sec:beatification},
by employing the infinite-order beatifying transformation, the precise relation between the original Casimir 
invariants, presented in section~\ref{sec:casimir}, and the corresponding beatified functionals was derived.
The Casimir invariants have many potential applications in the practical study of noncanonical Hamiltonian
systems. For example, the time-independent value of a Casimir functional can be used to validate a numerical 
solution.

A very common procedure in theoretical physics is the dimensional reduction of large dynamical systems with the 
purpose of enabling the application of numerical methods or generating low-dimensional models for the description 
of specific physical mechanisms. As already discussed in a previous work,\cite{Viscondi16b} a direct reduction 
in the degrees of freedom of a noncanonical Hamiltonian system can be quite problematic, as the truncated version 
of a noncanonical Poisson bracket generally does not satisfy the Jacobi identity. That is, excluding special and 
accidental cases, the Hamiltonian structure of a noncanonical system is eliminated by dimensional reduction. As 
a consequence, many fundamental properties of the dynamical system are possibly lost, such as the incompressibility 
of phase-space volumes, which prevents the occurrence of attractors.\cite{Morrison05} Beatification is a useful 
tool in preparing a noncanonical Hamiltonian system for proper dimensional reduction, since the Jacobi identity 
is preserved when a variable-independent Poisson operator undergoes a truncation process. 

Another relevant application of the beatification method is as an intermediate step toward the canonization 
of Hamiltonian systems. More than this, the preliminary use of the beatification procedure can be seen as 
a significant part of a systematic approach for obtaining canonical variables in complex Hamiltonian systems.
As such, beatification provides access to a wide array of analytical and numerical methods requiring a canonical 
Hamiltonian representation.


\section*{Acknowledgements}

This work was financially supported by FAPESP under grant numbers 2011/19296-1 and 2012/20452-0 
and by CNPq under grant numbers 402163/2012-5 and 470380/2012-8. In addition, PJM received support 
from DOE contract DE-FG02-04ER-54742 and funding provided by the Alexander von Humboldt Foundation. 


\appendix

\section{Beatification proof}
\label{app:brackettrans}

In this appendix, we present the proof that transformation~\eqref{eq:4.7} reduces 
expression~\eqref{eq:4.5} to the beatified Poisson operator~$\mcl{J}(\omega_{0})$,
as anticipated by equation~\eqref{eq:4.6}. As a first step toward this goal, 
we perform the first variation of identity~\eqref{eq:4.7}: 
\begin{equation}
 \delta\eta=\suml_{j=0}^{\infty}\frac{\vep^{j}}{j!}
 \mcl{B}^{j}\mu^{j}\delta\mu.
 \label{eq:A.1}
\end{equation}

According to the above equation, variations in the fields 
$\mu$ and $\eta$ are related by the following linear 
operator:
\begin{equation}
 \frac{\delta\eta}{\delta\mu}f=\suml_{j=0}^{\infty}\frac{\vep^{j}}{j!}{\mcl{B}}^{j}{\mu}^{j}f,
 \label{eq:A.2}
\end{equation}
\noindent where $f$ is an arbitrary function on the domain~$\mcl{D}$. In preparation 
for manipulating identity~\eqref{eq:4.5}, we now present some special properties of 
the operators $\mcl{J}$, ${\mcl{B}}$, and ${\mcl{B}}\dg$. First, as a direct consequence 
of equation~\eqref{eq:4.14}, note that the adjoint operator~${\mcl{B}}\dg$ satisfies the 
Leibniz's rule, that is, ${{\mcl{B}}\dg}fg=f{{\mcl{B}}\dg}g+g{{\mcl{B}}\dg}f$ for any two 
functions $f$ and $g$ on the domain~$\mcl{D}$. By performing successive applications of 
this rule, an important identity for the powers of the operator~${\mcl{B}}\dg$ is obtained:
\begin{equation}
 ({\mcl{B}}\dg)^{m}fg=f({\mcl{B}}\dg)^{m}g
 +\suml_{n=1}^{m}({\mcl{B}}\dg)^{m-n}({{\mcl{B}}\dg}f)({\mcl{B}}\dg)^{n-1}g.
 \label{eq:A.4}
\end{equation}

On account of its high relevance to our subsequent developments, we also present 
an interesting relation connecting the operators $\mcl{J}$, ${\mcl{B}}$, and 
${\mcl{B}}\dg$:
\begin{equation}
 {\mcl{B}}f\mcl{J}(\omega_{0})g=-\mcl{J}(f)g-\mcl{J}(\omega_{0})f{{\mcl{B}}\dg}g,
 \label{eq:A.5}
\end{equation}
\noindent which is valid for any functions $f$, $g$, and $\omega_{0}$ on the domain~$\mcl{D}$, 
under the condition that the operators ${\mcl{B}}$ and ${\mcl{B}}\dg$ are defined in 
terms of the state~$\omega_{0}$, in accordance with equations \eqref{eq:4.8} and \eqref{eq:4.14}. 
Notice that, in the particular case of a function~$f$ with constant value, identity~\eqref{eq:A.5}
reduces to ${\mcl{B}}\mcl{J}(\omega_{0})g=-\mcl{J}(\omega_{0}){{\mcl{B}}\dg}g$.

By employing equations \eqref{eq:A.2} and \eqref{eq:A.5}, we manipulate the following product 
of operators:
\begin{equation}
 \begin{aligned}
 \frac{\delta\eta}{\delta\mu}\mcl{J}(\omega_{0})
 =&\;\mcl{J}(\omega_{0})+\suml_{j=1}^{\infty}
 \frac{\vep^{j}}{j!}{\mcl{B}}^{j}{\mu}^{j}\mcl{J}(\omega_{0})\\
 =&\;\mcl{J}(\omega_{0})-\suml_{j=1}^{\infty}
 \frac{\vep^{j}}{j!}{\mcl{B}}^{j-1}[\mcl{J}({\mu}^{j})
 +\mcl{J}(\omega_{0}){\mu}^{j}{\mcl{B}}\dg]\\
 =&\;\mcl{J}(\omega_{0})-\suml_{j=1}^{\infty}
 \frac{\vep^{j}}{(j-1)!}{\mcl{B}}^{j-1}{\mu}^{j-1}\mcl{J}({\mu})\\
 &+\suml_{j=1}^{\infty}(-1)^{j}\frac{\vep^{j}}{j!}\mcl{J}(\omega_{0})
 ({\mcl{B}}\dg)^{j-1}{\mu}^{j}{{\mcl{B}}\dg},   
 \end{aligned}
 \label{eq:A.6}
\end{equation}
\noindent which is found on the right-hand side of equation~\eqref{eq:4.5}. In elaborating the above 
result, we have also made use of the identity $\mcl{J}({\mu}^{j})=j{\mu}^{j-1}\mcl{J}({\mu})$,
which directly follows from definition~\eqref{eq:2.2}. Upon substitution of expression~\eqref{eq:A.6}, 
the Poisson operator~\eqref{eq:4.5} is readily reformulated as follows:
\begin{equation}
 \til{\mcl{J}}=\mcl{J}(\omega_{0})\left[ 
 1+\suml_{j=1}^{\infty}(-1)^{j}
 \frac{\vep^{j}}{j!}({\mcl{B}}\dg)^{j-1}
 {\mu}^{j}{\mcl{B}}\dg\right]
 {\left(\frac{\delta\eta}{\delta\mu}\right)\!}\dg.
 \label{eq:A.7}
\end{equation}

By inserting the adjoint of equation~\eqref{eq:A.2} into the above identity, 
further manipulations are performed on the transformed Poisson operator:
\begin{equation}
 \begin{aligned}
 \til{\mcl{J}}=&\;\mcl{J}(\omega_{0})\left[1+\suml_{j=1}^{\infty}(-1)^{j}
 \frac{\vep^{j}}{j!}({\mcl{B}}\dg)^{j-1}{\mu}^{j}{\mcl{B}}\dg\right]
 \left[1+\suml_{j=1}^{\infty}\frac{\vep^{j}}{j!}{\mu}^{j}({\mcl{B}}\dg)^{j}\right]\\
 =&\;\mcl{J}(\omega_{0})\left[1+\suml_{j=1}^{\infty}(-1)^{j}
 \frac{\vep^{j}}{j!}({\mcl{B}}\dg)^{j-1}{\mu}^{j}{\mcl{B}}\dg
 +\suml_{j=1}^{\infty}\frac{\vep^{j}}{j!}{\mu}^{j}({\mcl{B}}\dg)^{j}\right.\\
 &\left.+\suml_{j,k=1}^{\infty}(-1)^{j}\frac{\vep^{j+k}}{j!k!}
 ({\mcl{B}}\dg)^{j-1}{\mu}^{j}{\mcl{B}}\dg{\mu}^{k}({\mcl{B}}\dg)^{k}\right]\\
 =&\;\mcl{J}(\omega_{0})
 +\mcl{J}(\omega_{0})\left\{\suml_{j=2}^{\infty}\frac{\vep^{j}}{j!}
 \left[(-1)^{j}({\mcl{B}}\dg)^{j-1}{\mu}^{j}+{\mu}^{j}({\mcl{B}}\dg)^{j-1}\right]\right.\\
 &\left.+\suml_{j=2}^{\infty}\suml_{k=1}^{j-1}(-1)^{j-k}\frac{\vep^{j}}{(j-k)!k!}
 ({\mcl{B}}\dg)^{j-k-1}{\mu}^{j-k}
 {{\mcl{B}}\dg}{\mu}^{k}({\mcl{B}}\dg)^{k-1}\right\}{{\mcl{B}}\dg}\\
 =&\;\mcl{J}(\omega_{0})+\mcl{J}(\omega_{0})\mcl{O}{{\mcl{B}}\dg},
 \end{aligned}
 \label{eq:A.8}
\end{equation}
\noindent where, for notational convenience, we have introduced the operator~$\mcl{O}$,
whose explicit expression is given within the curly brackets on the right-hand side of 
the third equality. For the purpose of simplifying equation~\eqref{eq:A.8}, we now make 
use of identity~\eqref{eq:A.4} in obtaining two new auxiliary relations:
\begin{subequations}
 \label{eq:A.9}
 \begin{align}
 &\,({\mcl{B}}\dg)^{j-1}{\mu}^{j}f={\mu}^{j}({\mcl{B}}\dg)^{j-1}f
 +j\suml_{k=1}^{j-1}({\mcl{B}}\dg)^{j-k-1}{\mu}^{j-1}
 ({{\mcl{B}}\dg}{\mu})({\mcl{B}}\dg)^{k-1}f,
 \label{eq:A.9a}\\
 &\begin{aligned}
 ({\mcl{B}}\dg)^{j-k-1}{\mu}^{j-k}{{\mcl{B}}\dg}{\mu}^{k}({\mcl{B}}\dg)^{k-1}f
 =&\;({\mcl{B}}\dg)^{j-k-1}{\mu}^{j}({\mcl{B}}\dg)^{k}f
 +k({\mcl{B}}\dg)^{j-k-1}{\mu}^{j-1}({{\mcl{B}}\dg}{\mu})({\mcl{B}}\dg)^{k-1}f\\
 =&\;{\mu}^{j}({\mcl{B}}\dg)^{j-1}f+k({\mcl{B}}\dg)^{j-k-1}
 {\mu}^{j-1}({{\mcl{B}}\dg}{\mu})({\mcl{B}}\dg)^{k-1}f\\
 &+j\suml_{m=k+1}^{j-1}({\mcl{B}}\dg)^{j-m-1}{\mu}^{j-1}
 ({{\mcl{B}}\dg}{\mu})({\mcl{B}}\dg)^{m-1}f,
 \end{aligned} 
 \label{eq:A.9b}
 \end{align}
\end{subequations}
\noindent in which $f$ is again an arbitrary function on the domain~$\mcl{D}$. In deriving 
the above equations, we have also used the identity~${\mcl{B}}\dg{\mu}^{j}=j{\mu}^{j-1}{\mcl{B}}\dg{\mu}$, 
which follows from the Leibniz's rule for the operator~${\mcl{B}}\dg$. By employing equations~\eqref{eq:A.9}, 
we recast the operator~$\mcl{O}$ in a more convenient form:
\begin{equation}
 \begin{aligned}
 \mcl{O}=&\;\suml_{j=2}^{\infty}\vep^{j}\left[\frac{(-1)^{j}+1}{j!}
 +(-1)^{j}\suml_{k=1}^{j-1}\frac{(-1)^{k}}{(j-k)!k!}\right]{\mu}^{j}({\mcl{B}}\dg)^{j-1}\\
 &+\suml_{j=2}^{\infty}(-\vep)^{j}\suml_{k=1}^{j-1}\left[\frac{1}{(j-1)!}+\frac{(-1)^{k}}{(j-k)!(k-1)!}\right]
 ({\mcl{B}}\dg)^{j-k-1}{\mu}^{j-1}({{\mcl{B}}\dg}{\mu})({\mcl{B}}\dg)^{k-1}\\
 &+\suml_{j=2}^{\infty}(-\vep)^{j}\suml_{k=1}^{j-1}\suml_{m=k+1}^{j-1}\frac{(-1)^{k}j}{(j-k)!k!}
 ({\mcl{B}}\dg)^{j-m-1}{\mu}^{j-1}({{\mcl{B}}\dg}{\mu})({\mcl{B}}\dg)^{m-1}.
 \end{aligned} 
 \label{eq:A.10}
\end{equation}

With the aid of the binomial theorem, the terms within square brackets in the first line 
of the above equation are shown to cancel one another. After making use of this fact, the 
operator~$\mcl{O}$ can be further simplified by switching the labels and the order of the 
summations over the indices $k$ and $m$ in the last line of equation~\eqref{eq:A.10}. In 
this way, we obtain:
\begin{equation}
 \mcl{O}=\suml_{j=2}^{\infty}(-\vep)^{j}\suml_{k=1}^{j-1}
 \left[\frac{(-1)^{k}}{(j-k)!(k-1)!}
 +\suml_{m=0}^{k-1}\frac{(-1)^{m}j}{(j-m)!m!}\right]
 ({\mcl{B}}\dg)^{j-k-1}{\mu}^{j-1}({{\mcl{B}}\dg}{\mu})({\mcl{B}}\dg)^{k-1}.
 \label{eq:A.11}
\end{equation}

The terms within square brackets in the above expression also cancel one another, as can be 
straightforwardly demonstrated by induction on the index~$k$. Therefore, identity~\eqref{eq:A.11} 
simply reduces to $\mcl{O}=0$. By substituting this result into equation~\eqref{eq:A.8}, we finally 
arrive at the defining relation~\eqref{eq:4.6}, thus concluding the proof that the beatification 
of the Poisson bracket~\eqref{eq:2.1} is provided by transformation~\eqref{eq:4.7}.



\end{document}